\documentclass[useAMS,usenatbib]{mn2e}

\input psfig.sty
\def\etal{et al.}

\title[The influence of halo assembly on galaxies and galaxy groups]{The influence of halo assembly on galaxies and galaxy groups}
\author[Zapata et al.]{Tatiana Zapata$^{1,2}$\thanks{E-mail:
taz@astro.iag.usp.br}, Josefa Perez$^{3}$, Nelson Padilla$^{1}$, Patricia Tissera$^{3}$\\
$^{1}$Departamento de Astronom\'\i a y Astrof\'\i sica, Pontificia Universidad Cat\'olica de Chile, Santiago, Chile.\\
$^{2}$Instituto de Astronomia, Geof\'\i sica e Ci\^encias Atmosf\'ericas, Universidade de S\~ao Paulo, Brasil.\\
$^{3}$Instituto de Astronom\'\i a y F\'\i sica del Espacio,Conicet-UBA, CC67, Suc.28,Ciudad de  Buenos Aires, Argentina.}
\begin{document}

\date{Accepted ????. Received ????; in original form ????}

\pagerange{\pageref{firstpage}--\pageref{lastpage}} \pubyear{2008}

\maketitle

\label{firstpage}

\begin{abstract}

In this paper, we study the variations of group galaxy properties according to 
the assembly history in SDSS-DR6 selected groups. Using mock SDSS group catalogues, 
we find two suitable indicators of group formation time: i) the  isolation of the group, defined as the
distance to the nearest neighbor in terms of its virial radius, 
and ii) the concentration, measured as the group inner density calculated using the fifth nearest bright galaxy
to the group centre.  Groups within narrow ranges of  mass in the mock catalogue show 
increasing group age with isolation and concentration.  
However, in the observational data the stellar age, as indicated by the spectral type, only shows a correlation
with concentration.
            
We study groups of similar mass and 
different assembly history, finding important
differences in their galaxy population.
Particularly, in high mass SDSS groups, the
number of members, mass-to-light ratios, red
galaxy fractions and the magnitude
difference between the brightest and second
brightest group galaxies, show 
different trends as a function of isolation and concentration,
even when it is expected that the latter two quantities correlate with group age.
Conversely, low
mass SDSS groups appear to be less sensitive
to their assembly history.

The correlations detected in the SDSS are not consistent with the
trends measured in the mock catalogues.
However, discrepancies can be
explained in terms of the disagreement found
in the age-isolation trends, suggesting that
the model might be overestimating the effects
of environment. We discuss how the modeling of
the cold gas in satellite galaxies 
could be responsible for this problem.
These results can be used to 
improve our understanding of the evolution of galaxies
in high-density environments.

\end{abstract}

\begin{keywords}
galaxies: clusters, galaxies: general, galaxies: surveys
\end{keywords}

\section{Introduction}

Up to only a few years back the mass of a dark-matter halo was the only 
parameter considered to affect its clustering properties in an important or detectable way (see for instance
Padilla \& Baugh 2002; Sheth, Mo \& Tormen 1998; Mo \& White 1996) as well as the galaxy population inside haloes.
The latter has been studied in the framework of the Halo Model (Cooray \& Sheth 2002; 
Cooray, 2005, 2006, and references therein),
whereby the number of galaxies per dark-matter halo,
as well
as central and satellite luminosities and possibly colours, are
proposed to depend only on the host-halo mass  (Wang et al., 2008); by assuming this, most
of the global fundamental statistics of the galaxy population can be then
reproduced.

Recent studies on the clustering of dark-matter haloes found that the age and assembly history
of haloes of similar mass influence their clustering amplitude.  The dependence on halo age
was first reported by Gao, Springel \& White (2005), and has been studied into more detail 
finding a general dependency on the way in which a halo is assembled by Gao \& White (2007),
Jing, Suto \& Mo (2007), Croton, Gao \& White (2007), and Wechsler et al., (2007).
The assembly bias has also been detected in observations by Wang et al. (2008); their studies
only concentrate on the clustering amplitude of different samples selected according to
group colour which they associate with group age.

Our aim is to detect the effects of halo assembly on the galaxy population.  Note that
even though there have been several attempts to study the
environmental dependence of galaxy properties in the Sloan Digital Sky Survey 
(as for instance in Blanton \& Berlind, 2007, and O'Mill, Padilla
\& Lambas, 2008),
and other galaxy catalogues (in the 2-degree Field Galaxy Redshift Survey by Gonz\'alez et al., 2005),
none of these works relate their results to the influence of halo assembly.  In addition to using group
colour as an indicator of halo age we will  explore other parameters 
as indicators such as the degree of dynamical relaxation of a group
given by its apparent projected
shape (rounder objects tend to be older, see Paz et al., 2006) or concentration (Wechsler et al. 2007,
Gao \& White 2007), 
luminosity difference between the brightest
and second brightest galaxies in the group (used to identify old, fossil groups, Mendes de Oliveira
\& Carrasco 2007; von Benda-Beckmann et al. 2007), and group isolation.

This paper is organised as follows, Section 2 presents the sets of observational data used in this
work as well as the mock catalogues constructed from semi-analytic galaxies; in Section 3 we explore
different indicators of group age, Section 4 shows the variations of galaxy properties as a function
of group isolation and concentration.  Section 5 presents a discussion
and Section 6 summarises the main results of this work.

\section{Data}

\begin{figure*}
\begin{picture}(430,440)
\put(-20,220){\psfig{file=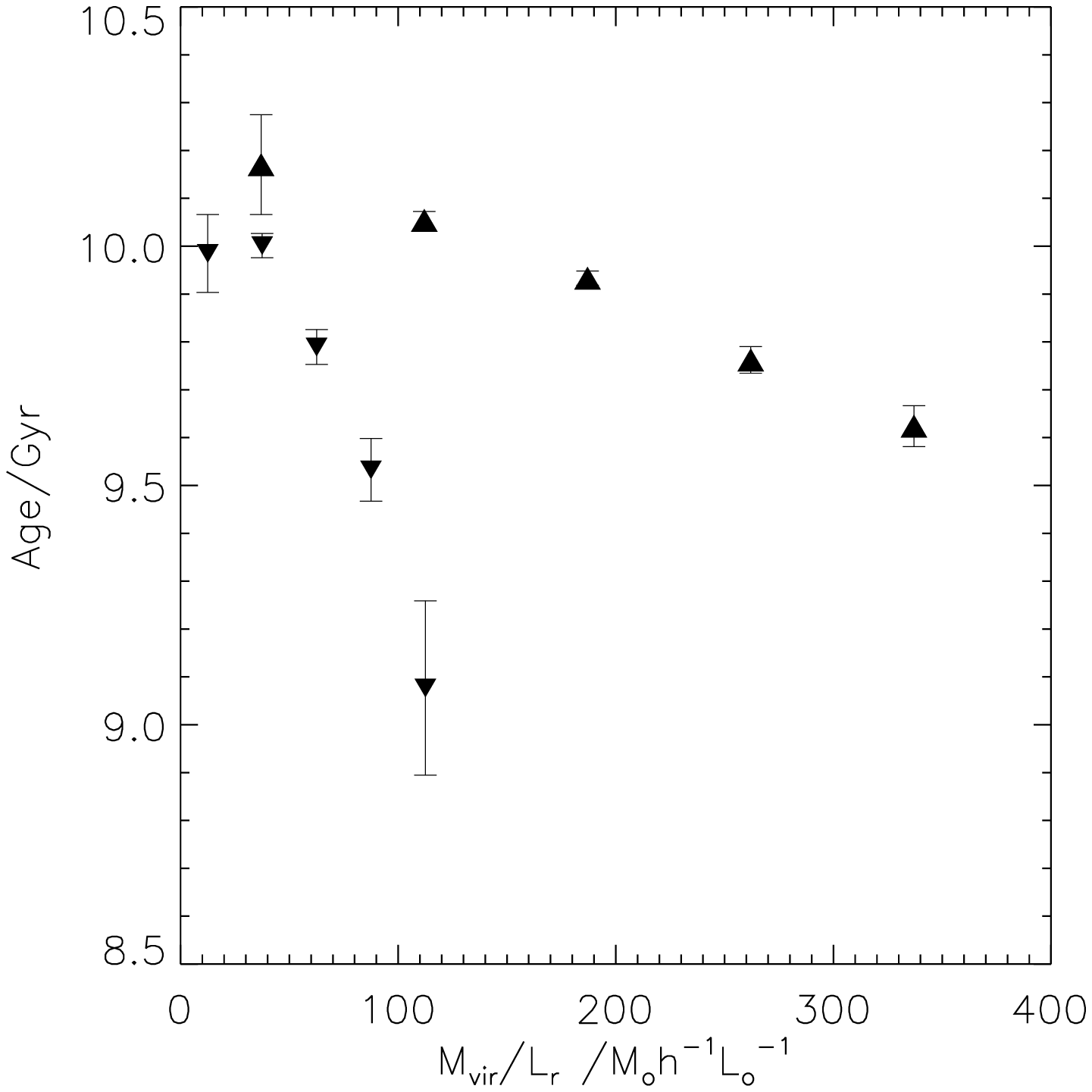,width=8.cm}}
\put(200,220){\psfig{file=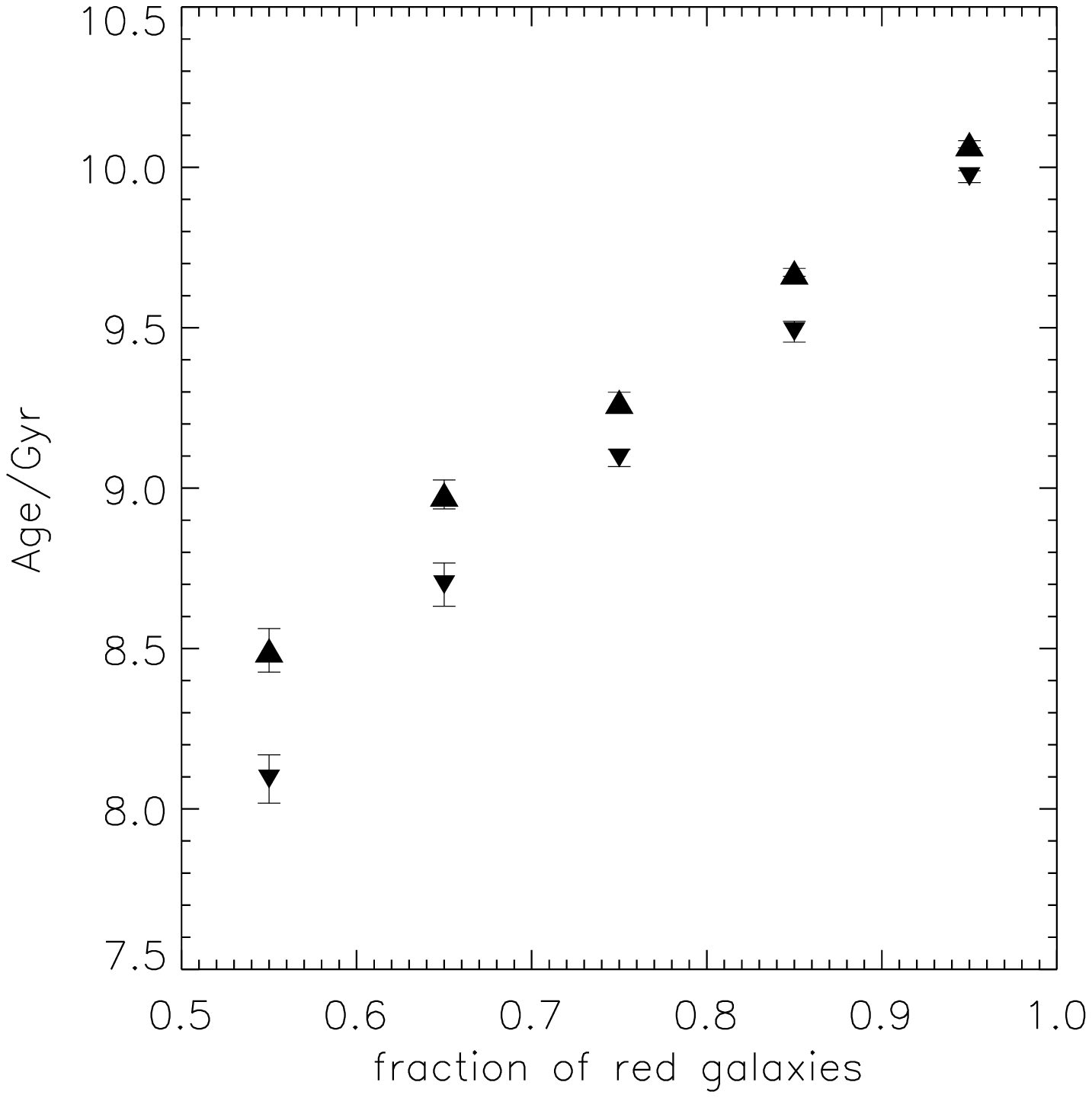,width=8.cm}}
\put(200,0){\psfig{file=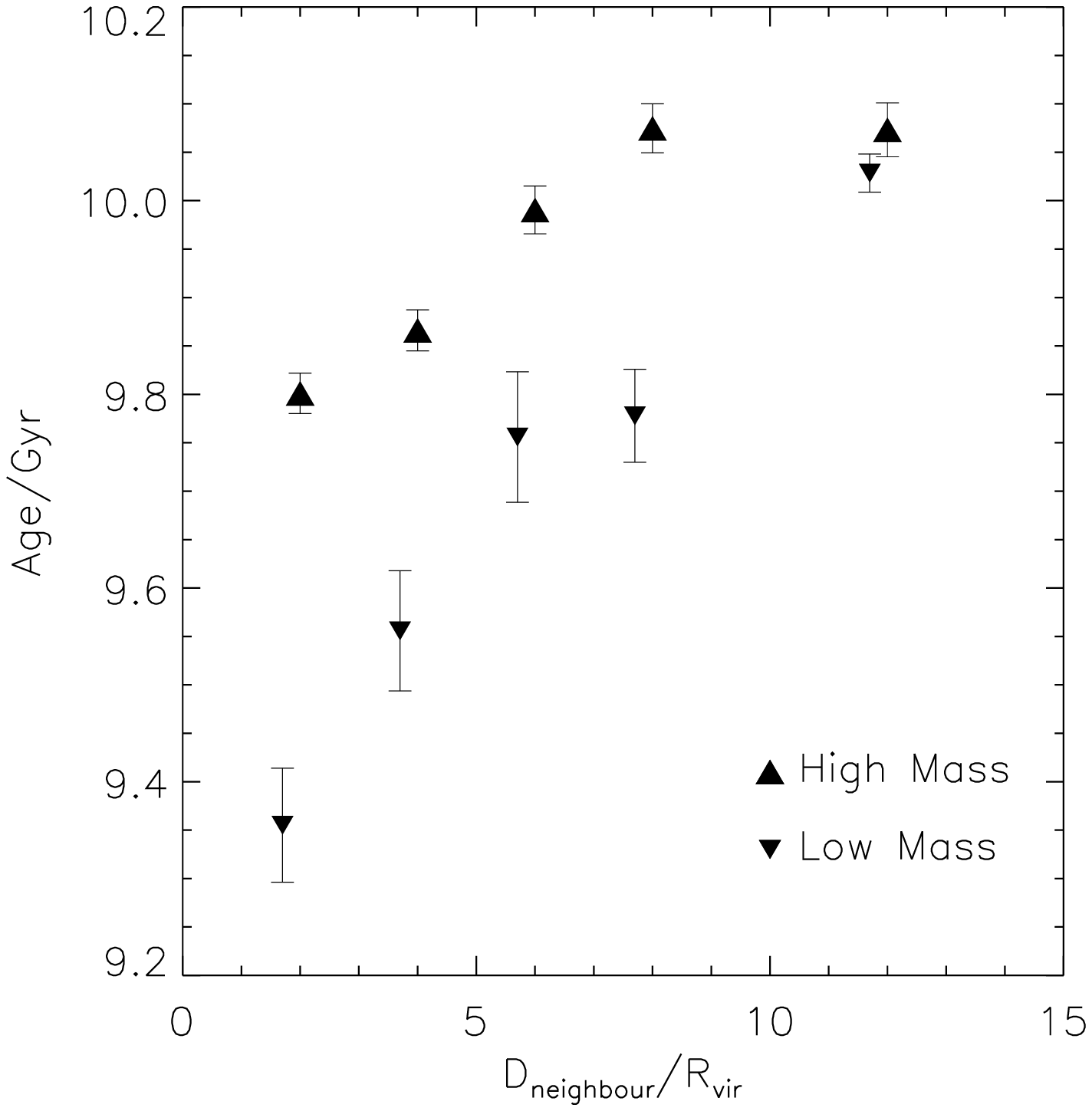,width=8.cm}}
\put(-20,0){\psfig{file=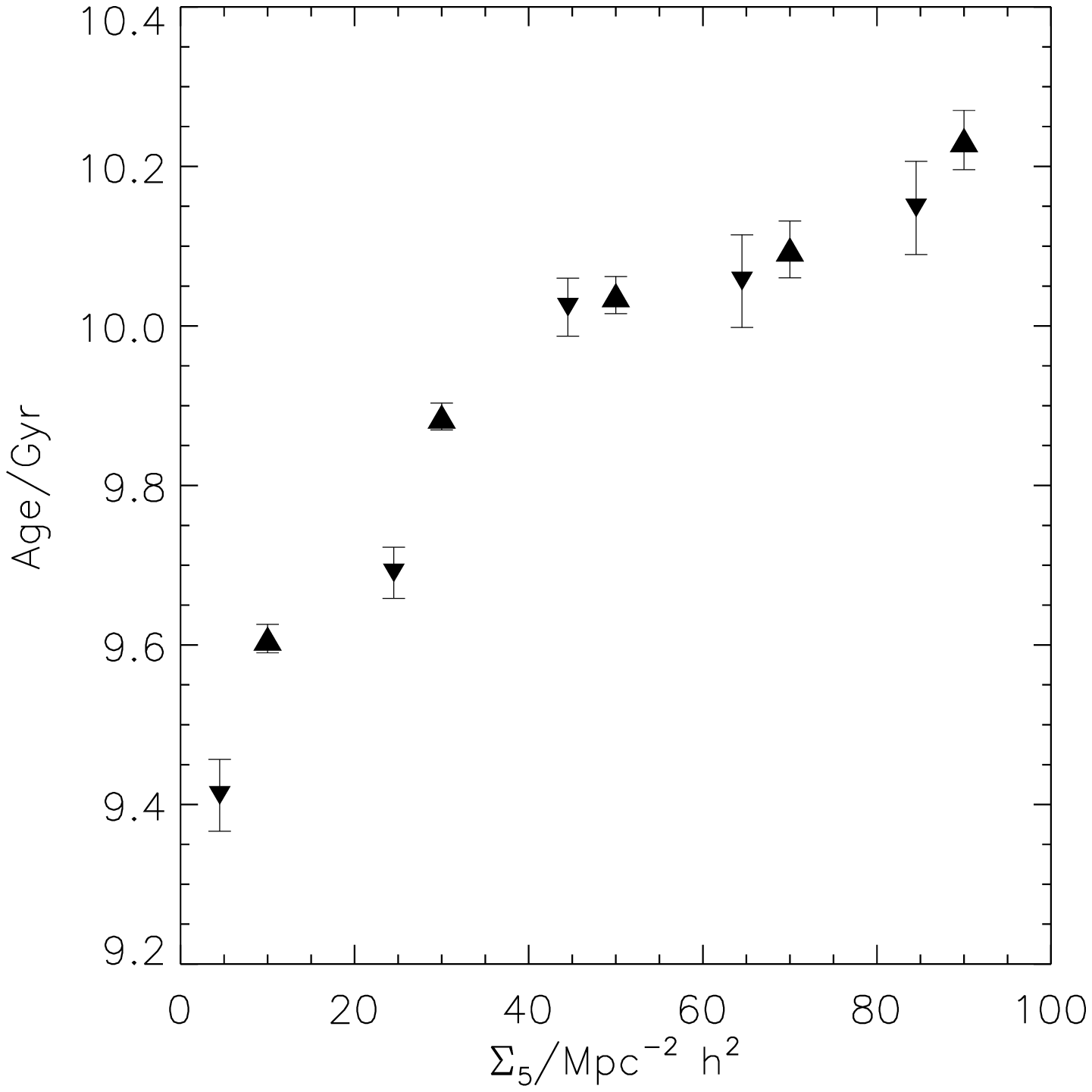,width=8.cm}}
\end{picture}
\caption{
Dependence of halo age on mass-to-light ratio (top left panel),
fraction of red galaxies (top right), on group inner density
using the five brightest galaxies in the group (bottom-left) and isolation
in terms of the group virial radius (bottom-right) in the Mock catalogues.
All panels show the relations 
for two different narrow (0.4 dex wide) 
ranges of halo mass; upward pointing triangles correspond 
to high masses, $13.6<\log_{10}(M_{\rm vir}$h$^{-1}$M$_{\odot})<14.0$
and downward triangles to low masses, $12.9<\log_{10}(M_{\rm vir}$h$^{-1}$M$_{\odot})<13.3$.
}
\label{fig:whoisage}
\end{figure*}

\subsection{Observations}
We use the Sloan Digital Sky Survey-Data Release 6 (SDSS-DR6, Adelman-McCarthy \etal, 2007) to identify groups following the prescription
presented by Merch\'an \& Zandivarez (2005), which consists on a friends-of-friends algorithm
with varying projected linking length $\sigma$, with $\sigma_0=0.239$ h$^{-1}$Mpc and fixed radial linking length 
$\Delta v=450 \ {\rm km s^{-1}}$.  These parameters correspond to the values found by Merch\'an \& Zandivarez to
produce a reasonably complete sample with low contamination ($95\%$ and $\la 8\%$, respectively).
{ The minimum number of galaxies per group in our sample is set to $10$.}

Even though there are previous compilations of groups in the SDSS such as the catalogues
compiled from the SDSS-DR3 (Merch\'an \& Zandivarez, 2005) and DR4 (Yang et al., 2007), and by using
percolation algorithms (Berlind et al., 2006) on the NYU-VAGC (Blanton et al. 2005), we need
to perform our own identification in order to control
the measurements of group centre, mass and membership.  Furthermore, given the low statistical significance
and high susceptibility to observational biases, we also need mock catalogues on which to perform
our analyses to test for these effects.  Therefore we apply exactly the same group detection algorithm to a SDSS mock
catalogue which we will describe in the following subsection.

We use the virial theorem to compute the virial mass of groups which is given by
\begin{equation}
 M_{vir}=\frac{3\sigma_{v}^2 R_{vir}}{G}
\end{equation}
where $\sigma_{v}$ is the line-of-sight velocity dispersion and $R_{vir}$ is estimated 
as in Merch\'an \& Zandivarez, (2005),
\begin{equation}
 R_{vir}=\frac{\pi}{2} \frac{N_{g}(N_{g}-1)}{\Sigma_{i>j}R_{ij}^{-1}}
\end{equation}
where $N_{g}$ is the number of galaxy members and $R_{ij}$ are the
galaxy relative projected distances.

{ Throughout this paper the
quoted errors in measurements involving SDSS groups will correspond to errors in the mean
obtained using the jacknife technique with $10$ subsamples.}

\subsection{Mock catalogues}

We construct a mock SDSS catalogue using the semi-analytic model (SAM) from De Lucia et al. (2006),
who use the merger trees from the Millennium Simulation (Springel et al., 2005) to follow different
physical processes that lead to the final $z=0$ galaxy population in a $\Lambda$CDM cosmology,  
 with  cosmological parameters determined from the combined analysis of the 2-degree
Field Galaxy Redshift Survey (2dFGRS, Colless et al. 2001)
and the first year Wilkinson Microwave Anisotropy Probe (WMAP) 
data (Spergel et al. 2003): $\Omega_{m}=0.25$, $\Omega_{b}=0.045$, 
$\Omega_{\Lambda}=0.75$, $h=0.73$, $n=1$ and $\sigma_{8}=0.9$. The
Millennium simulation consists of $2160^3$ particles, in a box of $500$h$^{-1}$Mpc on a side, with
a particle mass resolution of $8.6\times 10^8$h$ ^{-1}$M$_{\odot}$.  The resulting galaxies
conform a complete sample down to a magnitude $M_r=-16$.

In order to mimic the SDSS, we place four observers on different locations in the $z=0$ output
from the numerical simulation
so that the volumes surveyed by each observer are as independent as possible; this is achieved by
placing observers in the centre of four opposite octants in a diagonal direction, 
with two of the observers (on a long diagonal) facing in a direction
offset by 90 degrees (in the $x-y$ and $y-z$ planes) 
with respect to the other two observers.  The minimum distance between observers is 
$\simeq 350$h$^{-1}$Mpc, and they sample different regions of the simulation in different directions.
We apply the
same angular mask affecting the spectroscopic main sample, and apply a magnitude limit cut of 
$r_{\rm lim}=17.77$.  For each galaxy we store its observed properties, such as redshift (which
includes the peculiar motion), angular position in the sky, and apparent magnitudes in several bands.  However,
the main advantage of the mock catalogue is that we also store several intrinsic properties such as
luminosity in different bands and properties of the host halo including its mass.

We apply the adaptive FOF algorithm to each mock SDSS catalogue, using exactly the same
search parameters applied to the real data.  In this way, we obtain four mock group catalogues which
are studied the same way as the real SDSS groups; in particular, we use the same
{ minimum number of galaxies per group, as well as the same} method to estimate group masses.  

{ In the remainder of this paper the results obtained from the mocks will correspond to
the average over our four mock SDSS catalogues;
the uncertainties represented by the errorbars will correspond to the errors in the mean
obtained using the jacknife technique with $10$ subsamples, averaged over the mock catalogues.}

\section{Searching for observational indicators of group assembly} 

In this section, we analise the variation of  galaxy  group 
 properties with the aim of
finding an indicator of group age which can be related to the
 assembly history of the group host halo.
We will analyse several candidates of observational parameters that will 
allow us to rank our observational 
and simulated group samples according to their assembly history.

In the case of the mock catalogues, the group age is directly estimated by 
the mean stellar-mass weighted age of the galaxies hosted by a given group.
The star formation histories of galaxies in groups are determined by the rate of 
gas consumption which is modulated by the assembly history of the haloes (i.e. collapse, infall, mergers)
and feedback mechanisms which in effect are regulated by the size of the haloes.
A recent work by Li, Mo \& Gao (2008) shows that a halo age related to the formation history
of its member galaxies is the most appropriate for detecting the assembly effects
on the bias parameter; in this case the effect is seen for all halo masses, including
the ranges we will study in this paper,
$12.9<\log_{10}(M_{\rm vir}$ h$^{-1}$M$_{\odot})<14.0$.  This justifies our choice of halo age indicator.

Observationally, this parameter is more complicated to estimate. 
In order to find the best observational indicator which could be related 
to group age, we first search for possible correlations between 
the mock group ages and other group parameters.

\begin{figure*}
\begin{picture}(430,220)
\put(200,0){\psfig{file=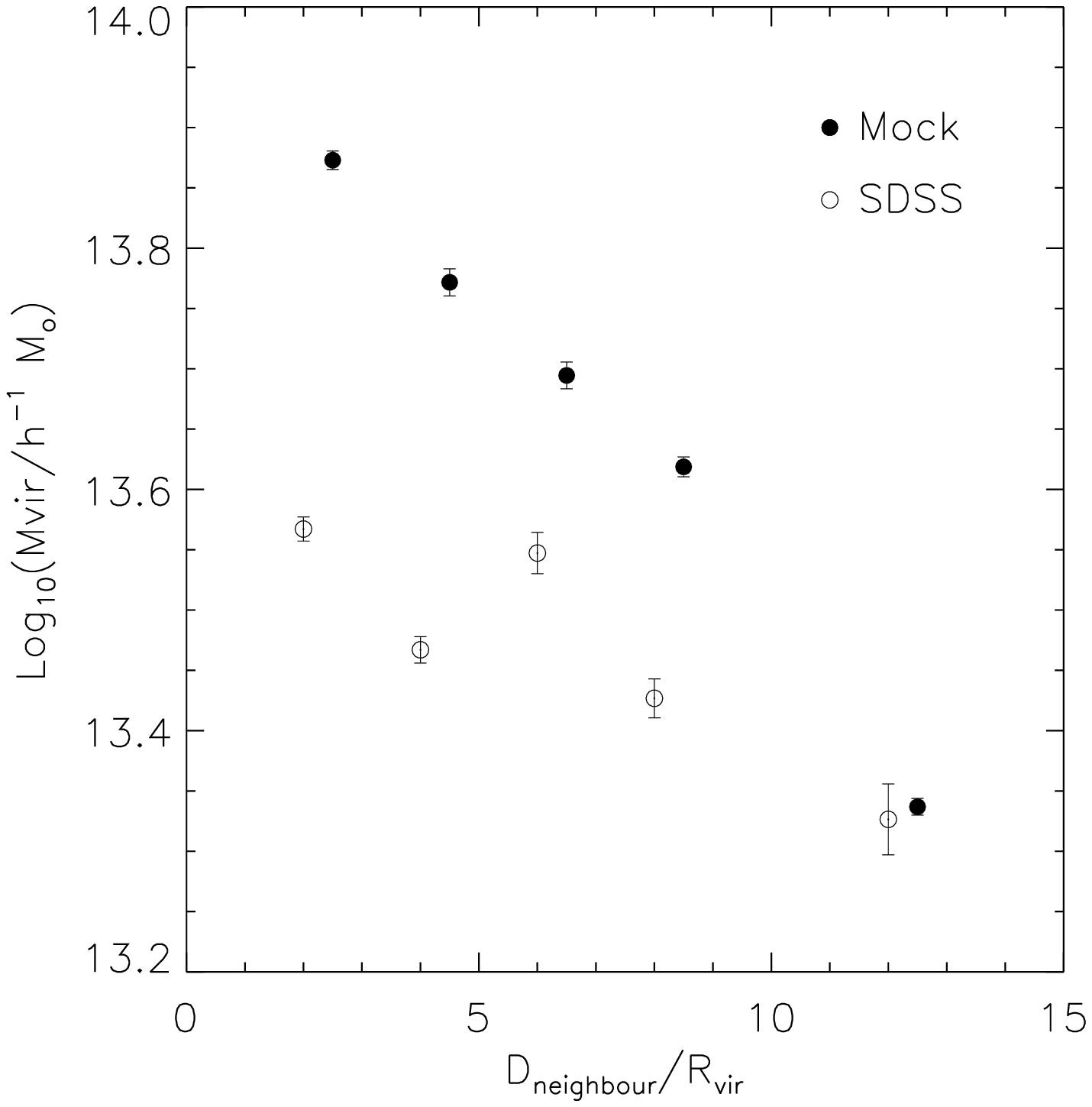,width=8.cm}}
\put(-20,0){\psfig{file=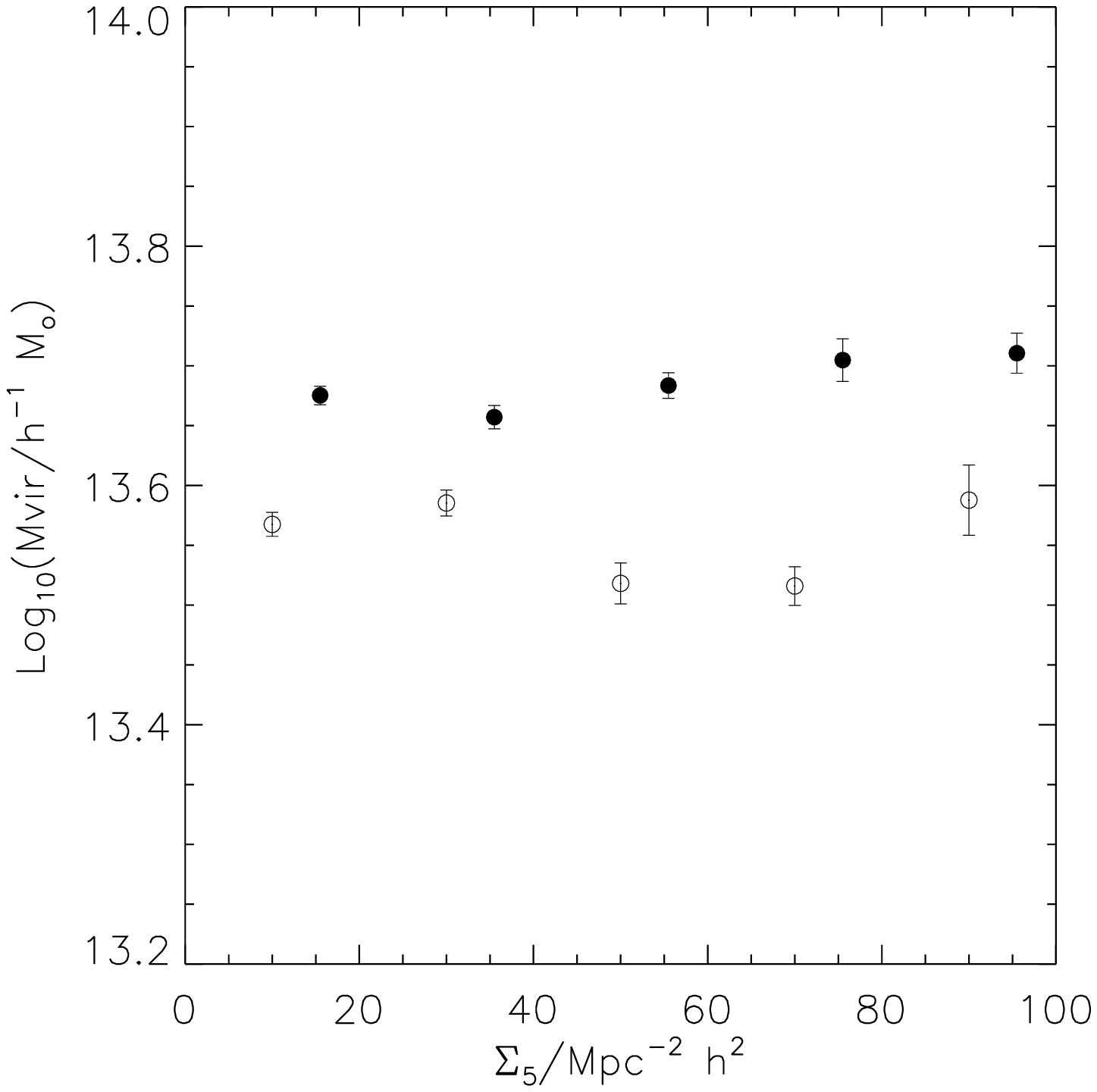,width=8.cm}}
\end{picture}
\caption{
Dependence of group mass on concentration (right) and
group isolation (left) for groups in the SDSS and mock catalogues.
}
\label{fig:mass}
\end{figure*}

\begin{figure*}
\begin{picture}(430,220)
\put(200,0){\psfig{file=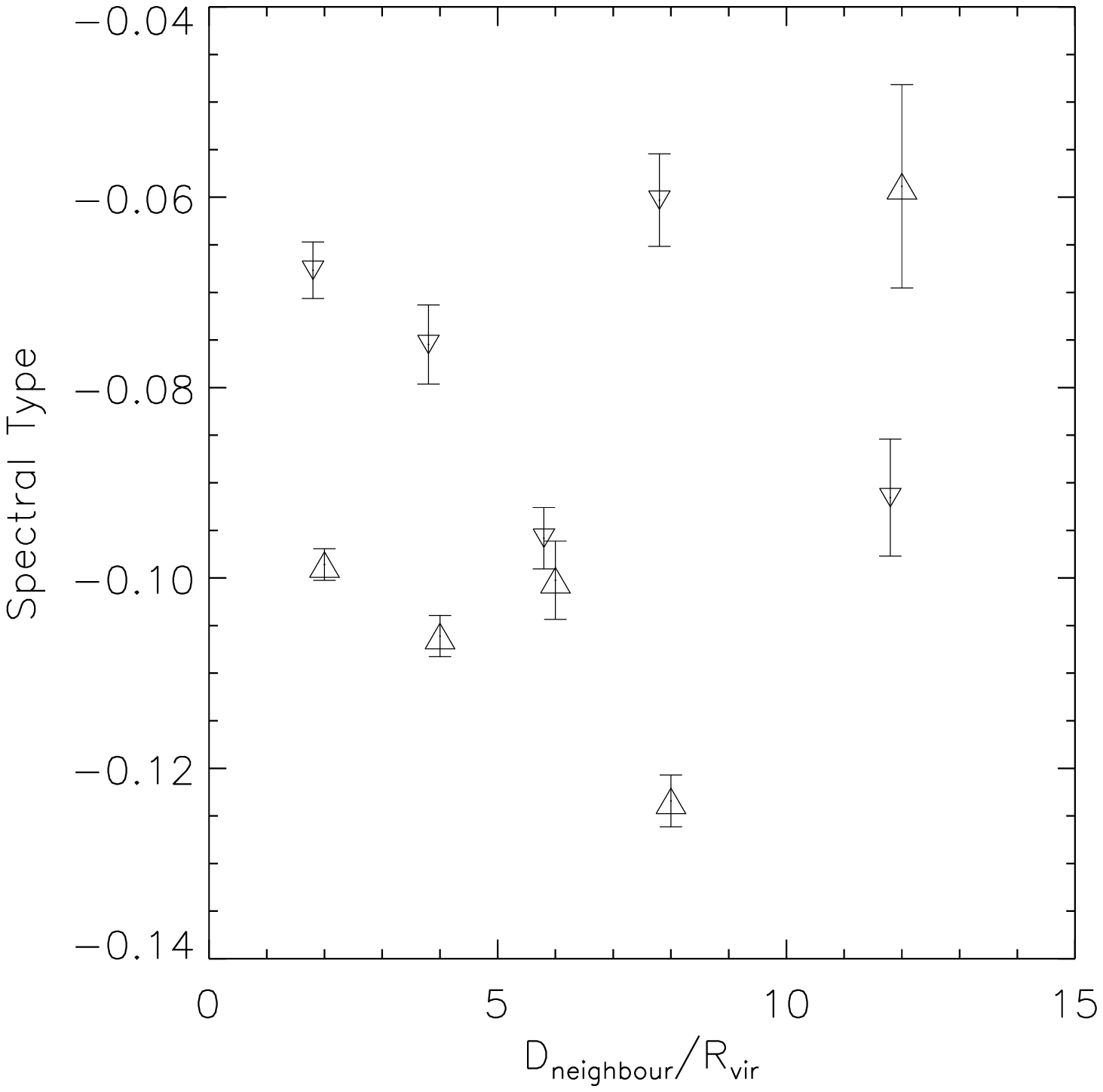,width=8.cm}}
\put(-20,0){\psfig{file=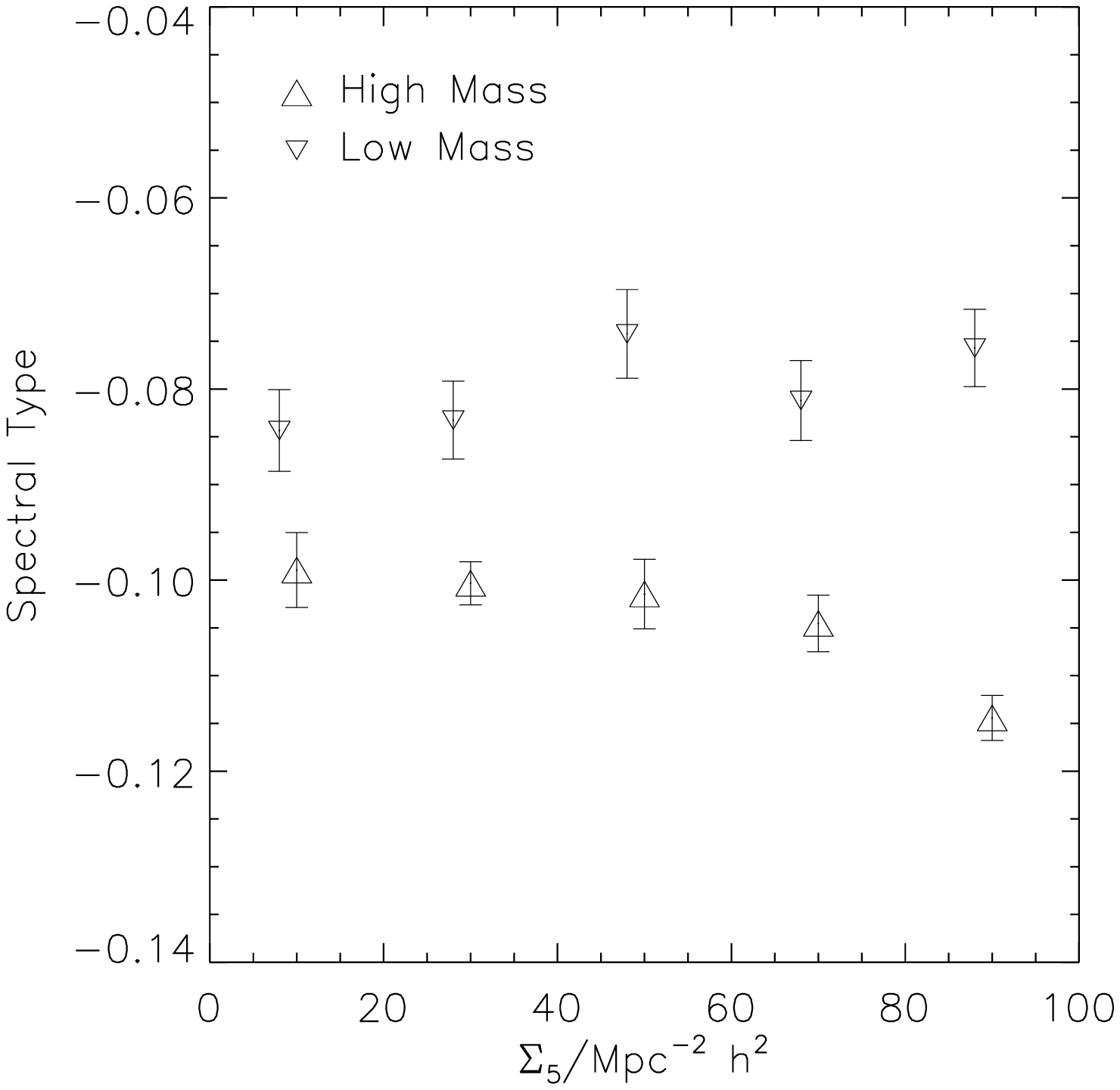,width=8.cm}}
\end{picture}
\caption{
Average group spectral type as a function of group concentration (left panel) and isolation (right panel) 
for groups in the SDSS separated according to the estimated virial mass as  high (upward pointing
triangles) and low (downward pointing triangles) group masses (see Fig.~\ref{fig:whoisage}).
}
\label{fig:edad}
\end{figure*}

Although the first candidate for age that comes to mind is the group mass 
as larger groups started to collapse earlier
(see for instance Paz et al. 2006),  our aim is to explore any possible 
variation of group properties with age for fixed values of group mass.
Hence, for fixed group masses, we test a series of observable parameters:

\begin{itemize}
\item the projected group
shape measured by the ratio between the minor and major semi-axes of an ellipse such that
the eigenvalues of the anisotropic part of its inertia tensor are the 
same as obtained from the projected
positions of the group members,
\item   the difference in magnitudes between the first and second
brightest galaxies in the group ($\Delta M_{12}$),
\item  the fraction of red galaxies in the group ($u-r \ge 2.0$),
\item   the degree of isolation of a group defined as
the distance $D_{\rm neighbor}/R_{\rm vir}$ (measured in terms of the group virial radius) to the nearest group neighbour,
\item the concentration measured by the group inner density ($\Sigma_5$), calculated 
by using the  projected distance to the 5$^{th}$ galaxy, brighter than M$_{\rm r} = -20.5 $,
and
\item  the mass-to-light ratio which we obtain using luminosities measured in the r-band.
\end{itemize}

We find that the group shape is not a good indicator of group age 
due to the lack of statistical significance of a relation between them.
Conversely, the fraction of red galaxies, the $M/L$ ratio, $D_{\rm neighbor}/R_{\rm vir}$ and $\Sigma_5$
show clear correlations with mean stellar-mass weighted ages
as it can be appreciated  in Fig.~\ref{fig:whoisage}.
In all cases we divide the group catalogues into subsamples 
defined by two narrow (0.4 dex wide) 
ranges of halo mass: high mass groups selected within 
 $13.6<\log_{10}(M_{\rm vir} $h$^{-1}$M$_{\odot})<14.0$
and low mass groups, within $12.9<\log_{10}(M_{\rm vir}$h$^{-1}$M$_{\odot})<13.3$ (from this point
on we will use the same halo mass ranges to study galaxy group properties as a function of different
age indicators).

Even though there is also a clear trend of age
with the fraction of red galaxies and the mass-to-light ratio, 
it has been shown that semi-analytic models overestimate the fraction of red galaxies
(e.g. Weinmann et al. 2006).  Conversely, the trends with group concentration  and isolation
are very well defined, showing that for the model, the most isolated and concentrated groups
tend to be older.   Since these parameters are 
more directly related to the structure of the dark matter, we adopt 
both estimators as good indicators of group age and assembly history.  
As it can be seen from the lower panels of Fig.~\ref{fig:whoisage}, a semi-analytic galaxy group with
no neighbour groups out to ten times its virial radius, or with an internal density larger than 
$80 \  h^2 {\rm Mpc^{-2}}$, is likely to be up to $800$ Myr older than
a group living in a more populated area or with a lower concentration.

This analysis also shows that both  group age indicators (isolation and  
concentration) depend on halo mass. In fact, 
in Fig.~\ref{fig:mass} we plot halo mass as a function of  concentration (left panel) and
isolation (right panel) for the mock (solid symbols) and the SDSS (open symbols).
Both the observations and model show a  trend with both parameters, however 
the dependence of mass on concentration  is  very weak,
indicating that the concentration may be better suited to
study assembly effects without introducing biases.
Nevertheless, given that the isolation of a group can store information on its large-scale 
neighbourhood, we will also include it in our analysis.

In order to detect dependencies of different group properties on these two age indicators,
we will separate groups  according to the proposed age parameter 
in two bins corresponding to young and old objects.  
Then for each bin, we calculate the mean group properties and errors using the jacknife technique, and
characterise the statistical significance of a trend by quoting the resulting 
difference between the galaxy population property in old and young groups ($\chi^2$).  
Notice that for one degree of freedom, $\chi^2=n$ indicates a $n\sigma$ detection.
In the case of concentration, we will consider as old objects those 
characterised by $\Sigma_5\geq70$h$^2$Mpc$^{-2}$, and
as young objects those with $\Sigma_5\leq30$h$^2$Mpc$^{-2}$.  
The limits for the case of isolation are set to
$D_{\rm neighbor}/R_{vir}\geq8$ for old objects and  $D_{\rm neighbor}/R_{vir}\leq7$ for young objects.
One important advantage of this approach is that correlations between different isolation or concentration
bins are reduced to a minimum.

A possible equivalent observational estimator of stellar mass age is the galaxy spectral
type (Yip et al. 2004); lower values of spectral type correspond
to earlier galaxy types and, therefore, to older stellar populations. In a similar
way, spectral types correlate with colours, corresponding to a value of -0.08
for the limit spectral type which separates the red and blue sequences. 
In order to test the validity of the previous theoretical indicators of age,
Fig. \ref{fig:edad} shows the mean spectral type of SDSS groups as a function 
of concentration (left panel) and isolation (right panel).  
For massive groups, we find different trends with these parameters.
While higher concentration massive SDSS groups are 
populated with systematically lower spectral type (i.e., older stellar populations, $4.85\sigma$ effect, 
see Table 1), the more isolated massive groups contain 
somewhat higher spectral type galaxies (i.e., younger, $1.39\sigma$ effect, see Table 1). 
These correlations 
show that SDSS groups confirm the predicted 
correlation between concentration
and age found in mock catalogues
(Fig.~\ref{fig:whoisage}). However a
disagreement in the SDSS and mock isolation-age/spectral type trends reveals that the model
may be overestimating the effects of environment. 
We will come back to this point in the following sections.
On the other hand, low mass SDSS groups show no significant trend (at most $1\sigma$).
with either isolation or concentration, again in contradiction to the findings 
obtained from the mock catalogues. 

In the following section, we turn to the study of the variation of galaxy
group properties as a function of assembly epoch as indicated by group concentration
and isolation.  When using isolation as an age indicator we
bear in mind the differences between observations and models.

\section{galaxy group properties as a function of group age}

\begin{table*}
 \centering
 \begin{minipage}{160mm}
  \caption{SDSS results: significance of trends in spectral type (column 3), mass-to-light
ratios (column 4), occupation numbers (column 5), fraction of red galaxies (column 6), and magnitude difference
between brightest and second brightest group member (column 7), as a function of concentration (top rows) and isolation
(bottom rows).  All quantities are rounded to $\sim1$ significant figure in the error, estimated using the jacknife technique.}
   \label{table1}
  \begin{tabular}{@{}ccccccc@{}}
  \hline
  Mass range & quantity & Spec. Type & $M_{vir}/L_r/(M_{\odot}$h$^{-1}L_{\odot}^{-1})$ & $Log_{10}(N_{gal})$ & fr. red gals. & $\Delta M_{12}$\\
  \hline
  high mass& low $\Sigma_5$                      &$-0.097\pm0.002$  &$572\pm13$    & $1.178\pm0.032$ &$0.784\pm0.004$  &$0.49\pm0.04$\\
           & high $\Sigma_5$                     &$-0.108\pm0.001$  &$443\pm8$     & $1.454\pm0.053$ &$0.809\pm0.005$  &$0.54\pm0.04$\\
           & $\chi^2$ (trend)                    &$4.85$ (decr.)    &$8.37$ (decr.)& $4.47$ (incr.)  &$4.02$ (incr.)   &$0.9$ (none)\\
  low mass & low $\Sigma_5$                      &$-0.08\pm0.002$   &$180\pm3$     & $1.065\pm0.017$ &$0.77\pm0.01$    &$0.49\pm0.06$\\
           & high $\Sigma_5$                     &$-0.077\pm0.002$  &$158\pm5$     & $1.064\pm0.013$ &$0.815\pm0.006$  &$0.64\pm0.04$\\
           & $\chi^2$ (trend)                    &$1.07$ (incr.)    &$3.94$ (decr.)& $0.006$(none)   &$3.84$ (incr.)   &$1.99$ (incr.)\\
  \hline
  high mass& low $D_{\rm neighbor}/R_{\rm vir}$  &$-0.099\pm0.001$  &$498\pm13$    &$1.302\pm0.021$ &$0.809\pm0.005$   &$0.45\pm0.01$\\
           & high $D_{\rm neighbor}/R_{\rm vir}$ &$-0.091\pm0.006$  &$704\pm40$    &$1.121\pm0.038$ &$0.763\pm0.016$   &$0.51\pm0.05$\\
           & $\chi^2$ (trend)                    &$1.39$ (incr.)    &$4.92$ (incr.)&$7.9$ (decr.)   &$2.7$ (decr.)     &$1.18$ (incr.)\\
  low mass & low $D_{\rm neighbor}/R_{\rm vir}$  &$-0.0735\pm0.001$ &$227\pm6$     &$1.086\pm0.014$ &$0.797\pm0.004$   &$0.44\pm0.02$\\
           & high $D_{\rm neighbor}/R_{\rm vir}$ &$-0.0733\pm0.003$ &$144\pm3$     &$1.069\pm0.027$ &$0.749\pm0.007$   &$0.52\pm0.02$\\
           & $\chi^2$ (trend)                    &$0.06$ (none)     &$11.51$ (decr.)&$8.02$ (decr.)  &$5.33$ (decr.)   &$2.53$ (incr.)\\
  \hline
  \end{tabular}
  \end{minipage}
\end{table*}

\begin{table*}
 \centering
 \begin{minipage}{160mm}
  \caption{Mock results: columns and rows are as in Table 1.
}
   \label{table2}
  \begin{tabular}{@{}ccccccc@{}}
  \hline
  Mass range & quantity & Spec. Type & $M_{vir}/L_r/(M_{\odot}$h$^{-1}L_{\odot}^{-1})$ & $Log_{10}(N_{gal})$ & fr. red gals. & $\Delta M_{12}$\\
  \hline
  high mass& low $\Sigma_5$                      &N/A &$202\pm2$     & $1.231\pm0.006$ &$0.863\pm0.002$  &$0.70\pm0.01$\\
           & high $\Sigma_5$                     &N/A &$177\pm2$     & $1.379\pm0.013$ &$0.913\pm0.002$  &$0.77\pm0.03$\\
           & $\chi^2$ (trend)                    &N/A &$10.3$ (decr.)& $10.0$ (incr.)  &$17.6$ (incr.)   &$2.23$ (incr.)\\
  low mass & low $\Sigma_5$                      &N/A &$60\pm1.6$    & $1.088\pm0.008$ &$0.854\pm0.005$  &$0.61\pm0.02$\\
           & high $\Sigma_5$                     &N/A &$56\pm1.7$    & $1.087\pm0.014$ &$0.938\pm0.006$  &$0.67\pm0.03$\\
           & $\chi^2$ (trend)                    &N/A &$1.9$ (decr.) & $0.09$ (none)   &$11.4$ (incr.)   &$1.45$ (incr.)\\
  \hline
  high mass& low $D_{\rm neighbor}/R_{\rm vir}$  &N/A& $198\pm2$     & $1.313\pm0.006$ &$0.881\pm0.002$  &$0.75\pm0.01$\\
           & high $D_{\rm neighbor}/R_{\rm vir}$ &N/A& $177\pm2$     & $1.227\pm0.009$ &$0.921\pm0.002$  &$0.73\pm0.02$\\
           & $\chi^2$ (trend)                    &N/A& $7.2$ (decr.) & $8.1$ (decr.)   &$11.28$ (incr.)  &$0.86$ (none)\\
  low mass & low $D_{\rm neighbor}/R_{\rm vir}$  &N/A& $64\pm1.1$    & $1.062\pm0.006$ &$0.852\pm0.01$   &$0.64\pm0.02$\\
           & high $D_{\rm neighbor}/R_{\rm vir}$ &N/A& $53\pm0.7$    & $1.07\pm0.01$   &$0.916\pm0.002$  &$0.64\pm0.02$\\
           & $\chi^2$ (trend)                    &N/A& $8.7$ (decr.) & $0.67$ (none)   &$8.4$ (incr.)    &$0.08$ (none)\\
  \hline
  \end{tabular}
  \end{minipage}
\end{table*}

In principle, halo models (cf. Cooray \& Sheth, 2002) and 
the conditional luminosity function (Cooray 2005, 2006)
are able to establish the properties of galaxies in dark-matter haloes using only the halo mass.  This
technique has been very successful in reproducing several galaxy statistics such as the correlation
function of SDSS galaxies (Zehavi et al., 2004) and numerical simulations (Zheng et al., 2005), and
the evolution of the luminosity function of galaxies up to $z=6$ (Cooray 2005).  However, as we will show in this
section, the properties of galaxies in groups of similar masses are also dependent on the 
concentration and isolation of the group,  parameters related to the assembly history of the system.

\begin{figure*}
\begin{picture}(430,220)
\put(200,0){\psfig{file=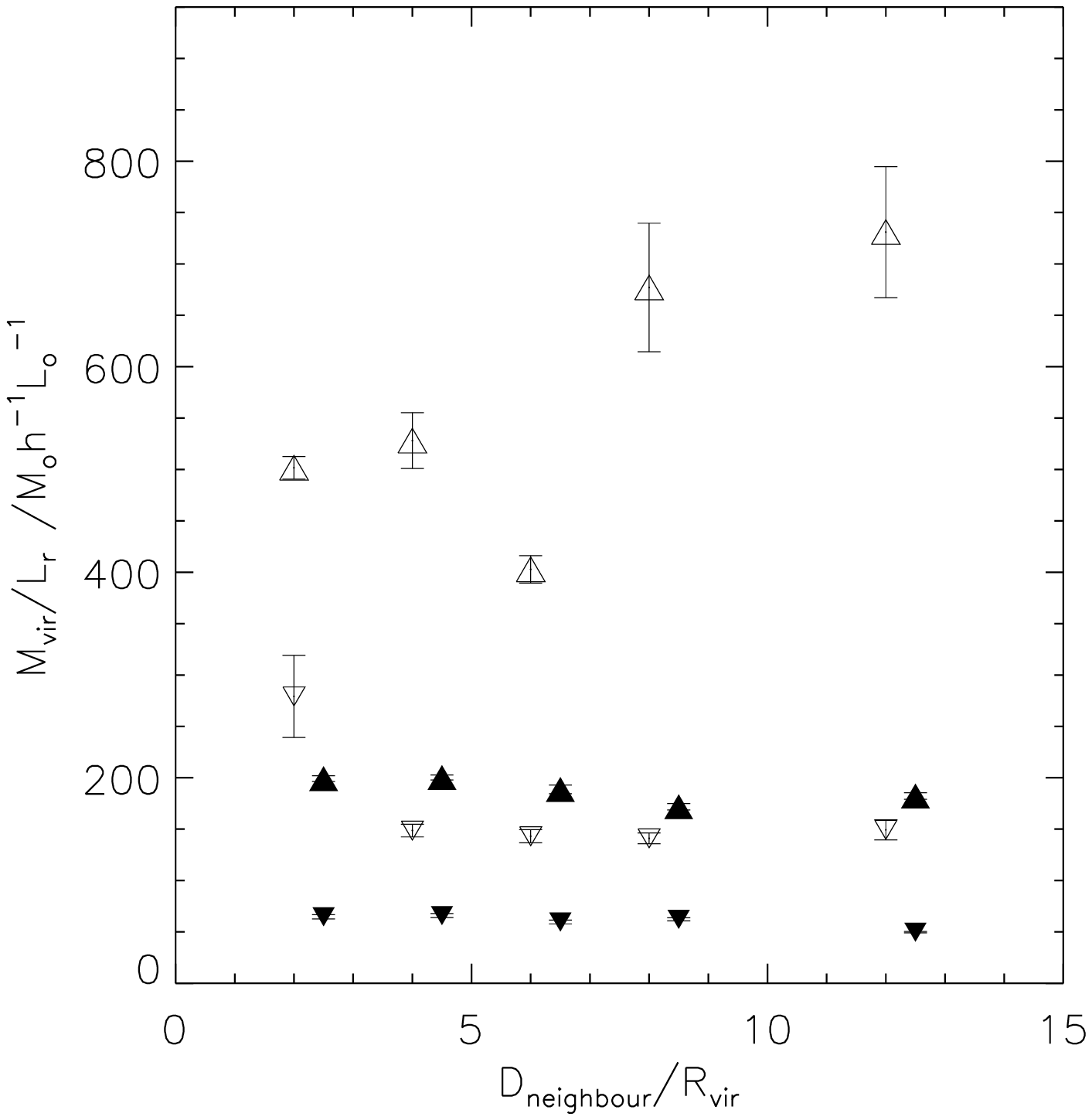,width=8.cm}}
\put(-20,0){\psfig{file=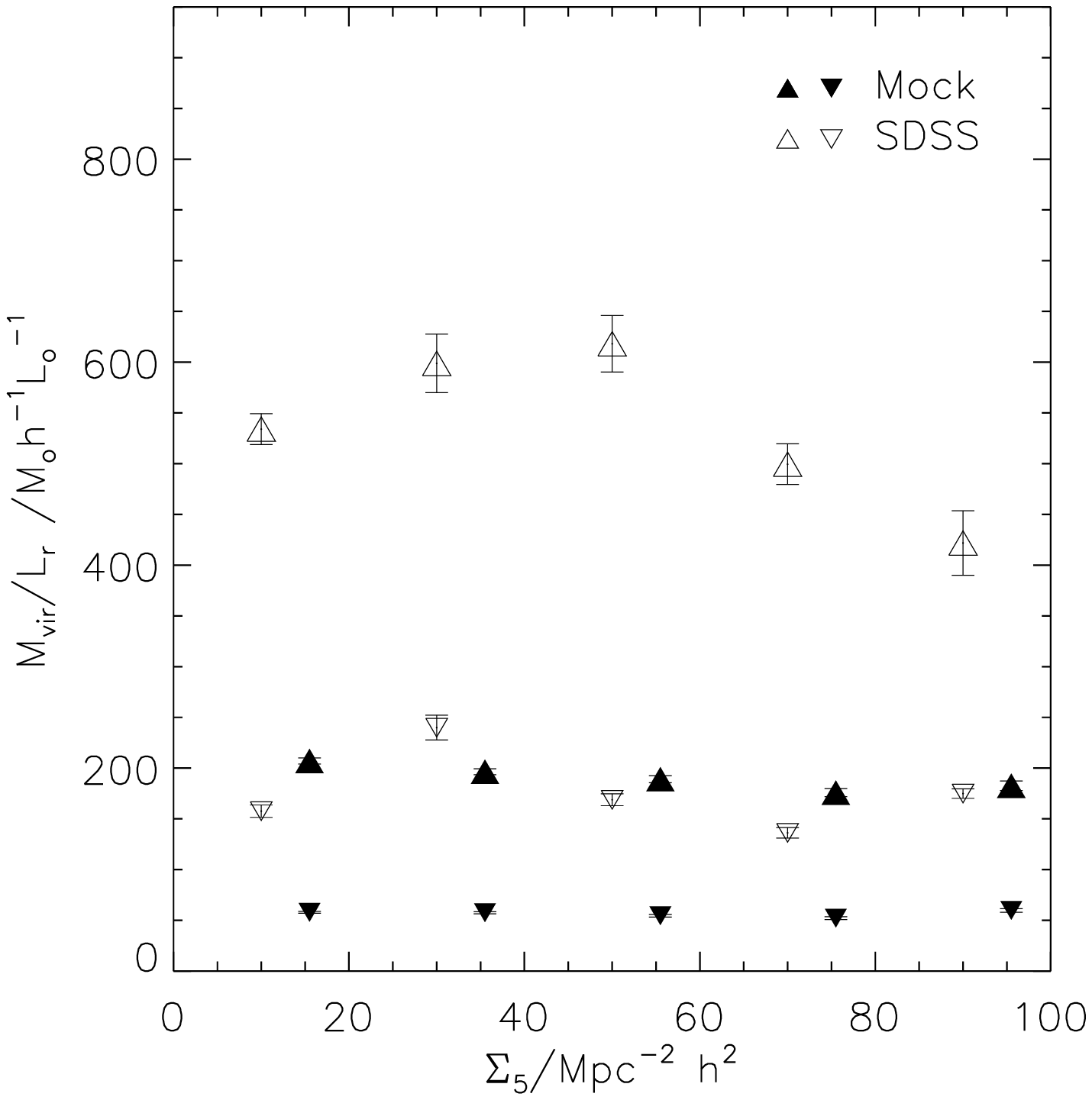,width=8.cm}}
\end{picture}
\caption{
Dependence of mass-to-light ratios on group concentration (left) and
isolation (right panel) for mock and SDSS (filled and open symbols,
respectively) groups, in two narrow
ranges of halo mass (high masses are shown in upward pointing triangles, low masses in downward triangles).
}
\label{fig:ml}
\end{figure*}

We start analysing the variations of mass-to-light ratios, $M/L$, of low and high mass groups
for different degrees of isolation and concentration.
Fig. ~\ref{fig:ml} shows the $M/L$ ratios as a function of group concentration
(left panel) and isolation (right panel).  
In consistency with our previous result for spectral types, high mass SDSS groups 
show two clear opposite trends of decreasing ($8.37\sigma$, Table 1)
and increasing $M/L$ ($4.92\sigma$, Table 1) with concentration and isolation, respectively, 
of almost a $50\%$ change. As can be seen, both trends are comparatively significant.
A more detailed analysis shows that $(90\pm10)\%$ of the $M/L$ ratios exhibited
by groups isolated by more than $8$ virial radii lies above the median relation between 
$M/L$ ratio to $\Sigma_5$; however, the values of $\Sigma_5$ 
for these isolated groups span almost the entire range of concentrations shown in the figure.
This tells us that more isolated groups at a given  concentration show 
higher $M/L$ ratios.
Note also that since the analysis is carried out in a narrow mass range, the 
increase or decrease in  $M/L$ can be directly related to a decrease or increase, respectively
in the r-band luminosity. Therefore, the opposite behaviours in the $M/L$ ratios observed for 
the SDSS high mass groups with concentration and isolation can also be associated to opposing
trends in their r-band luminosities. This can be explained 
in terms of their spectral types; while high mass SDSS groups tend to be
older with increasing concentration (and consequently brighter in the r-band),
they tend to be younger with increasing isolation (and fainter in the same band).

\begin{figure*}
\begin{picture}(430,220)
\put(200,0){\psfig{file=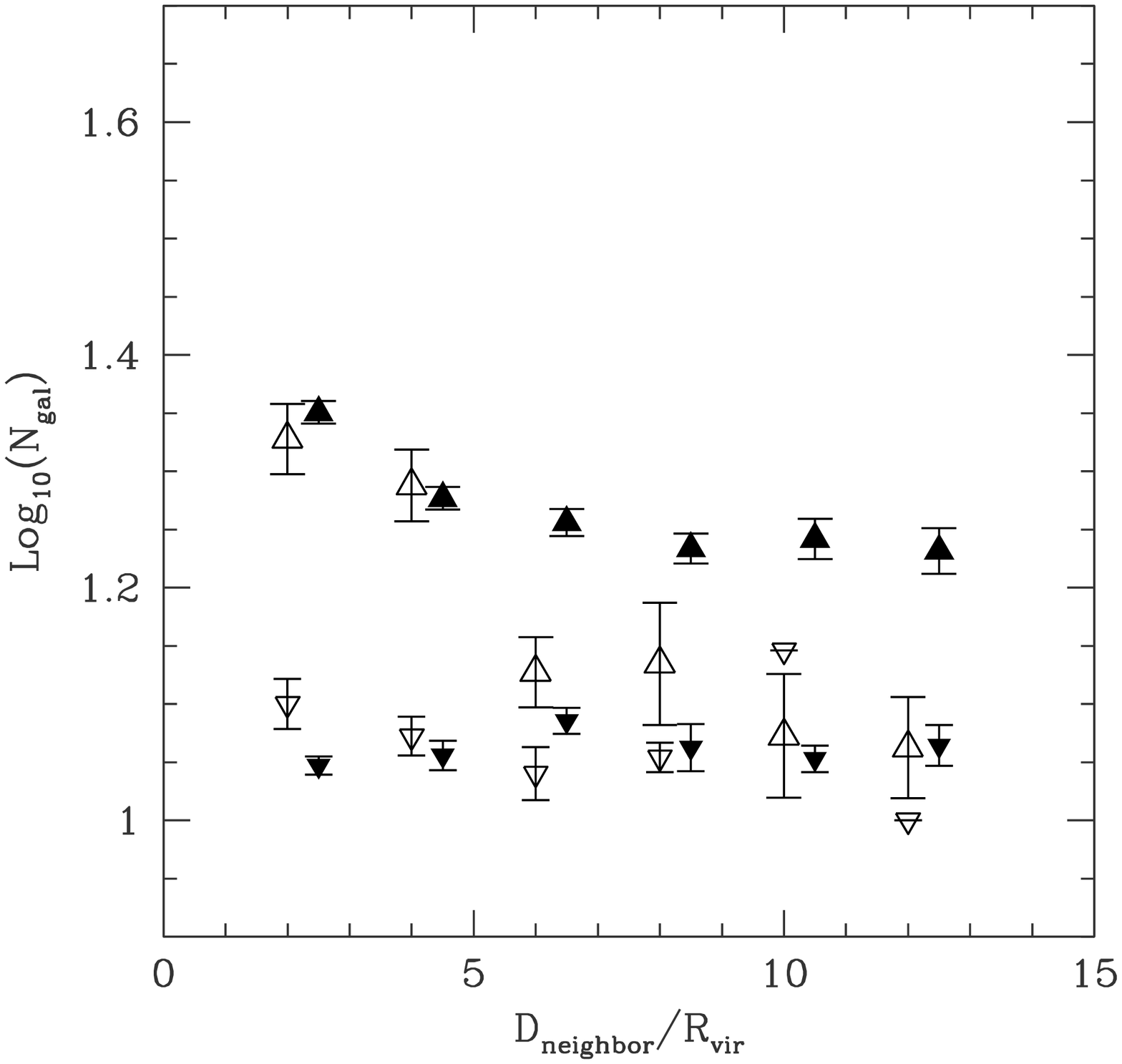,width=8.cm}}
\put(-20,0){\psfig{file=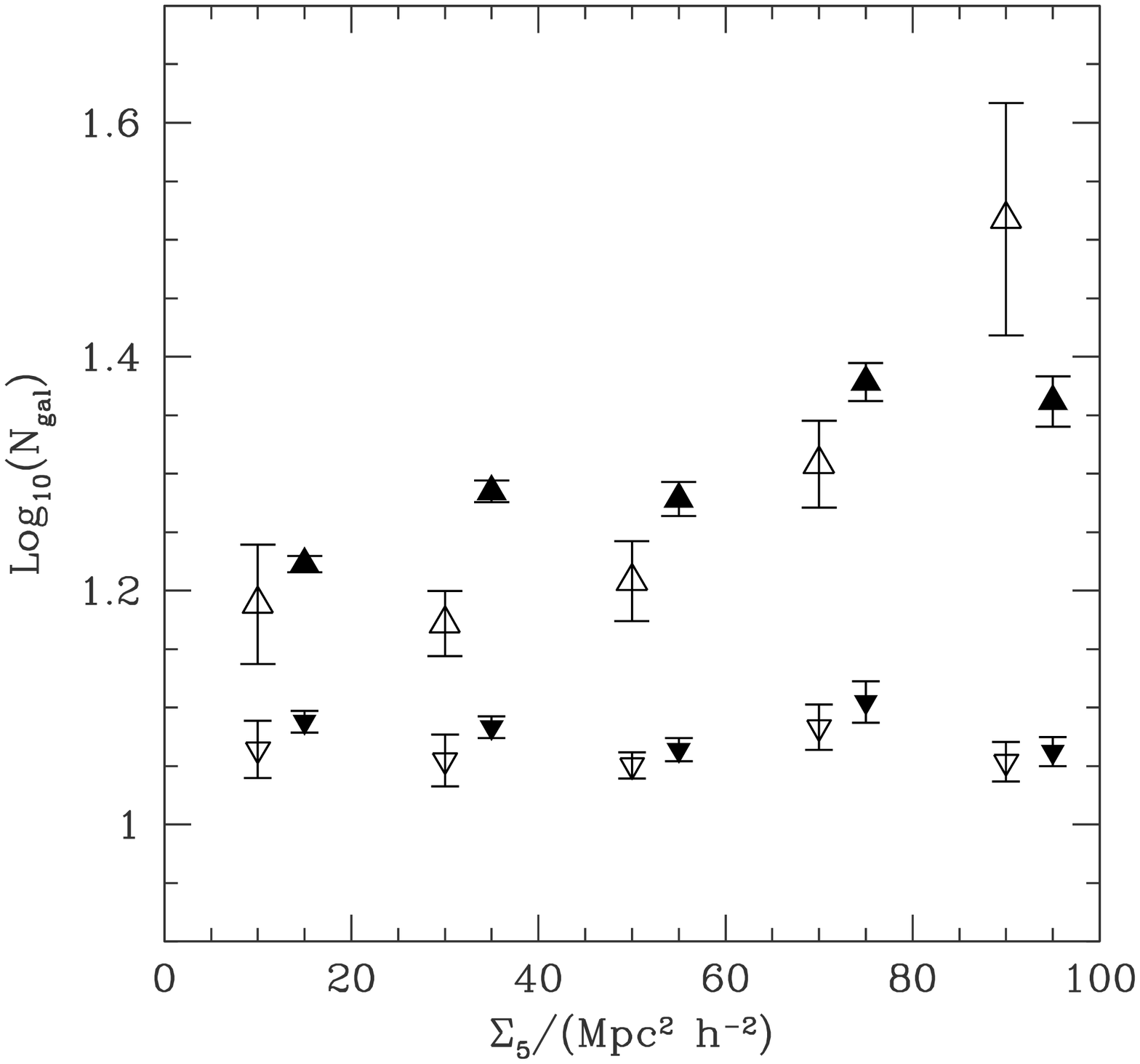,width=8.cm}}
\end{picture}
\caption{
Dependence of the number of galaxy members on group concentration (left) and
isolation (right panel) for mock and SDSS-DR6 groups (filled and open symbols,
respectively), in two narrow
ranges of halo mass (high masses are shown in upward pointing triangles, low masses in downward triangles).
}
\label{fig:numbers}
\end{figure*}

On the other hand, 
Fig. ~\ref{fig:ml} also shows that low mass SDSS groups
exhibit compatible trends of decreasing $M/L$ with both concentration and isolation, with
a high statistical significance (Table 1).
Interestingly, groups in the mock catalogues show this trend for both high 
and low mass groups (the mass-to-light ratios decreases
as a function of isolation and concentration).
This discrepancy between observational and mock catalogues for high mass groups is 
an indication of possible problems in the modeling of the
physical processes associated to the baryonic component in the SAM, 
since the mass-to-light ratio, $M/L$, is strongly dependent on the galaxy population.

In order to investigate further the origin of these behaviours,
we also estimate the number of group members  as a function of concentration and isolation, for both 
SDSS and mock groups. As it can be seen in Fig. \ref{fig:numbers}, 
the number of members for high mass SDSS groups decreases with isolation ($7.9\sigma$, Table 1)
and increases with concentration ($4.47\sigma$, Table 1). This result is also consistent 
with the $M/L$ and r-band luminosity tendencies since a decreasing number
of members produces lower total luminosities as the group isolation increases, and
an increasing number of galaxy members results in higher luminosities 
as the concentration increases.
Note that, although  high mass groups in the mock catalogues exhibit similar trends than the observations 
(see Table 2)
for the number of galaxy members as a function of concentration and isolation (Fig~\ref{fig:numbers}), 
they totally disagree in reproducing the $M/L$ behavior.  
This suggests that the intrinsic luminosity of the galaxy members vary differently in the model
and observations as a function of isolation and concentration.  
 This discrepancy between mock and SDSS groups 
may be an indication that the model fails in describing the 
physical processes associated to the baryonic component to some level. Several facts could be
involved: there could be an offset in the gas accretion rates onto the halo,
 the star formation timescales could be too short, the threshold cold gas surface density  for
 star formation could be too low, the suppression of star formation by AGN 
heating could be overestimated, the merging timescales could be too short, environmental
effects could be exacerbated, and so on. 
Because the most significant discrepancy is observed for the most massive haloes
in low density environments (large isolation), we can rule out some of these sources of 
disagreement. 
In the perspective of the current hierarchical model of structure formation,
where massive haloes assemble rather late through the merging of smaller systems,
gas cooling processes are the main contributors to the galaxy mass-growth in low mass haloes. 
On the other hand, galaxy-galaxy mergers 
are more efficient within small haloes with low velocity dispersion. 
As a consequence, both processes would not explain the observed disagreement. 
AGN feedback acts regulating the star formation activity in massive galaxies and 
has been proved to be reasonably well suited to reproduce several observational statistics, 
so it is not likely to be responsible for the present problem.
Thus, the remaining possible causes could lie in 
the star formation prescription or in environmental processes. 
Recently, analytic and hydrodynamic simulations have shown that the environment
in groups and cluster is less aggressive than previously assumed. It has
been argued that the excess of red satellite  owes to an oversimplified treatment of
ram pressure stripping of their hot gaseous halos. In general, SAMs assume that the entire hot gas reservoir
of a galaxy is instantaneously stripped at the moment of accretion into a larger halo 
(when it becomes a satellite galaxy). Recent improvements in SAMs allow
a better agreement with observations for the fractions of red satellites 
once the following considerations are implemented; (i) a decreased stripping efficiency,
and (ii) a significant loss of low-mass satellites tidally disrupted during the accretion by
a central host galaxy (Font et al 2008; Kang et al. 2008). Thus,
satellite galaxies are able to retain a significant fraction of their
hot gas for several Gyr., replenishing their reservoirs of cold, star forming gas and
remaining blue for relatively longer periods of time.

\begin{figure*}
\begin{picture}(430,220)
\put(-20,0){\psfig{file=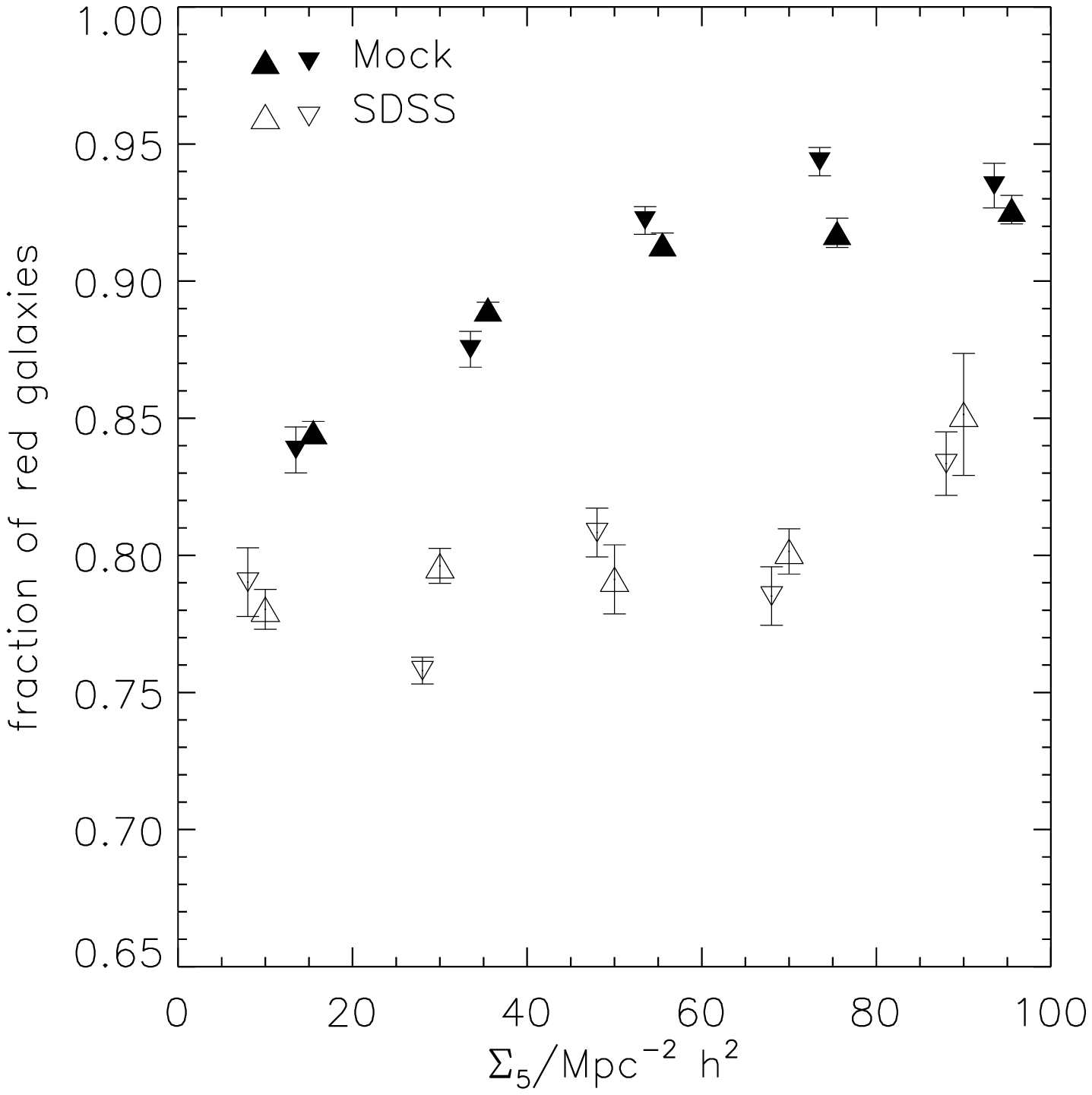,width=8.cm}}
\put(200,0){\psfig{file=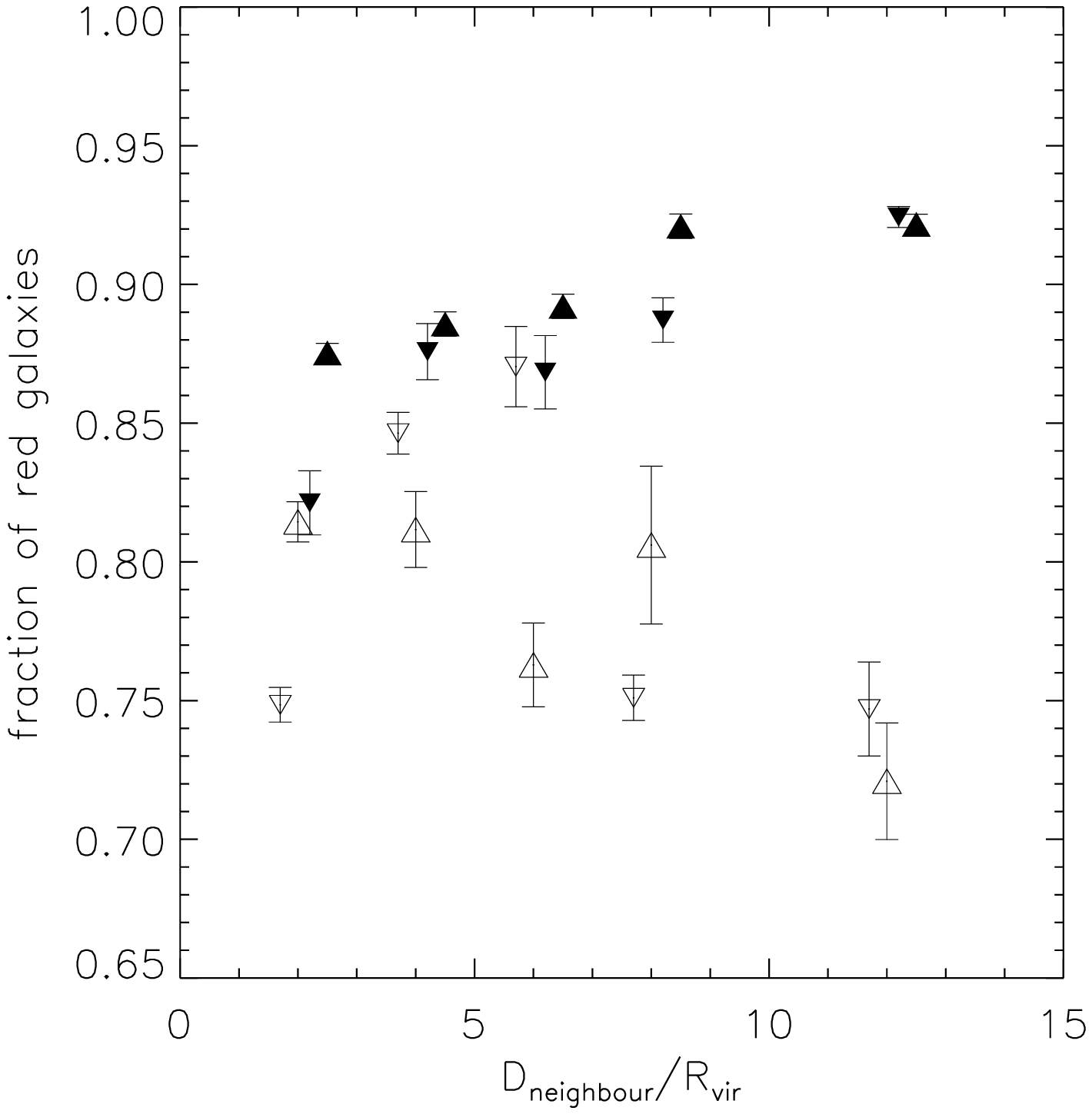,width=8.cm}}
\end{picture}
\caption{
Dependence of the red galaxy fraction on concentration (left panel) and isolation (right panel) for mock
(filled symbols) and SDSS (open symbols) groups in two narrow
ranges of mass (low masses indicated by downward pointing triangles, high masses by upward triangles).
}
\label{fig:frojas}
\end{figure*}

\begin{figure*}
\begin{picture}(430,220)
\put(-20,0){\psfig{file=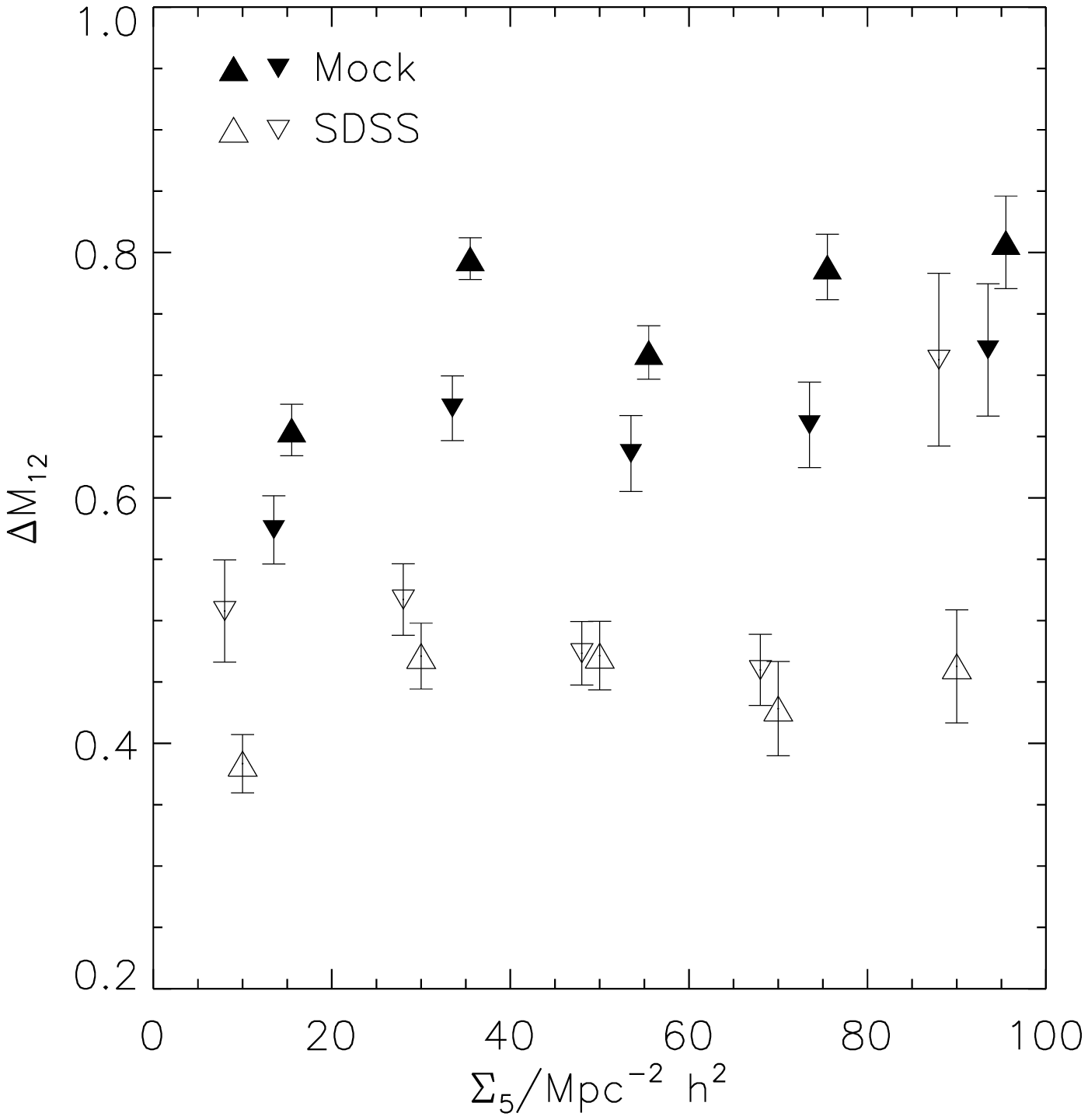,width=8.cm}}
\put(200,0){\psfig{file=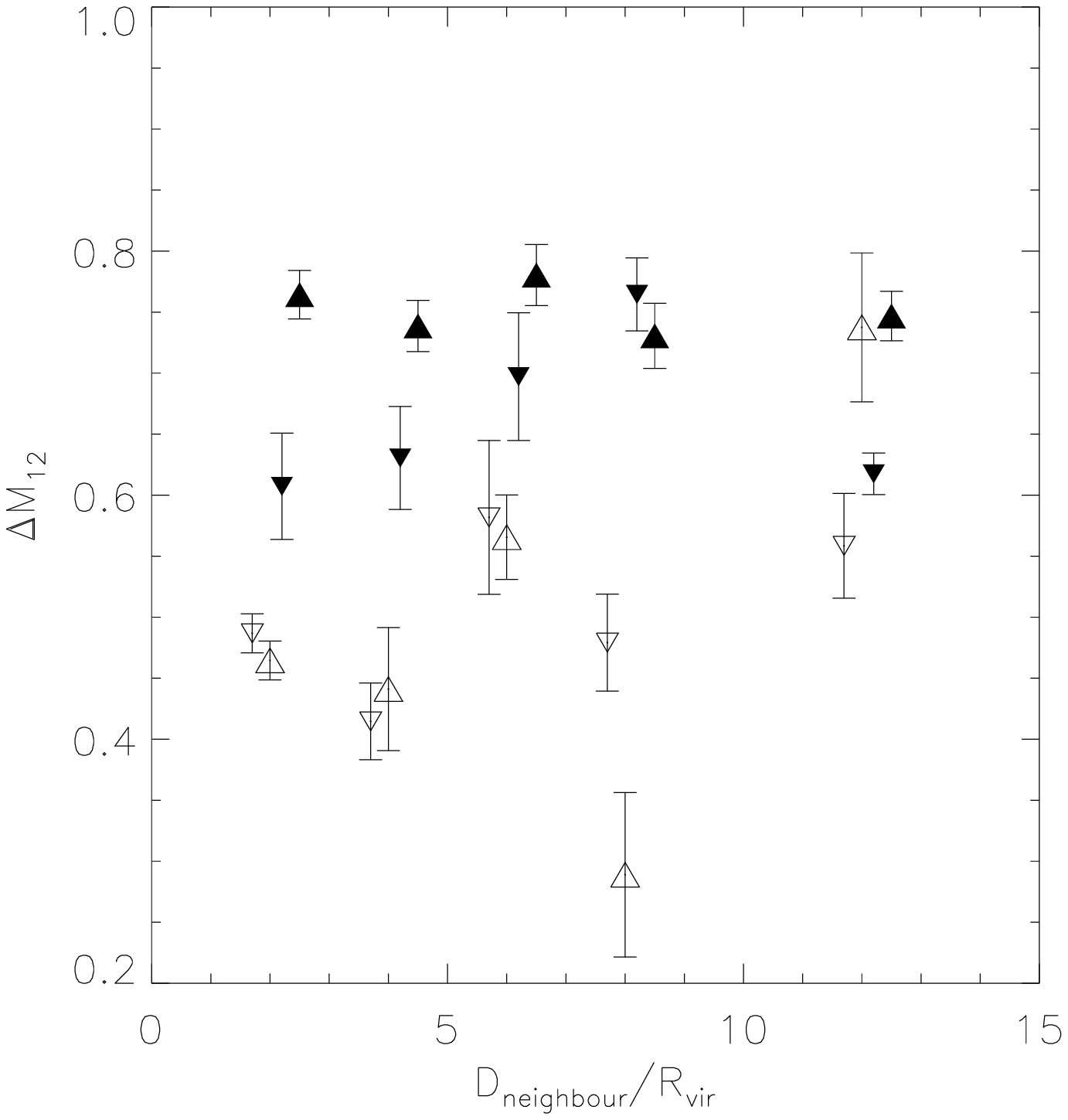,width=8.cm}}
\end{picture}
\caption{
Dependence of $\Delta M_{12}$ on group concentration (left panel) and isolation (right panel)
 for the mock (filled symbols) and
SDSS (open symbols) group catalogues. Upward pointing triangles refer to group masses 
$13.6<\log_{10}(M_{\rm vir} $h$^{-1}$M$_{\odot})<14.0$,
whereas downward pointing triangles to $12.9<\log_{10}(M_{\rm rvir}$h$^{-1}$M$_{\odot})<13.3$.
}
\label{fig:m12}
\end{figure*}

Given that the fraction of red galaxies is  a good indicator of the evolutionary state 
of a group (Bell et al. 2004),  we also investigate  the variations of this parameter
with concentration and isolation in the mock and   SDSS groups (Fig.\ref{fig:frojas}).  
We find that both SDSS and mock groups show a similar trend of increasing 
red galaxy fraction as a function of concentration  (left panel of the figure,
significance levels greater than $4\sigma$, Tables 1 and 2).  This result is
consistent with the expected behaviour since concentration correlates with age for both SDSS and mock groups. 
However, it is noticeable the different fractions
present in the model and in the SDSS; the former are $\sim 10\%$ larger,
in agreement with previous studies that indicate that semi-analytic colours in the
Croton et al. (2006) model are systematically redder than SDSS galaxies (Weinmann et al. 2006)\footnote{
 Croton \& Farrar (2008) study galaxy colours in voids in the 2-degree Field Galaxy Redshift
Survey (Colles et al., 2001), and find that the semi-analytic model by Croton et al. (2006)
produces similar galaxy colours in such low-density environments.  Notice that this
is not to be compared to our results since
in our case, even in the most isolated environment galaxies are embedded in virialised groups
of at least $10$ galaxy members, i.e., not in galaxy voids.
}.
On the other hand, as can be seen in the right panel of this figure, 
high mass SDSS groups show a decreasing fraction of red galaxies as  groups become more isolated,  
 in agreement with the larger fraction of late type galaxies found to populate them (Fig. \ref{fig:edad}).
We calculate how often extremely isolated SDSS massive groups with distances to their closest neighbors 
larger than $>8R_{\rm vir}$, show red galaxy fractions below the median as 
a function of $\Sigma_5$, and find that this occurs 
with a frequency of a $(70\pm17)\%$, again with $\Sigma_5$ values over the whole range shown 
in the figure.  This is in agreement with the results found 
for the $M/L$ dependence on isolation and concentration.

In contrast, as we have already discussed, mock catalogues
show that high mass groups have a larger fraction of red galaxies as the isolation 
increases. In fact, this disagreement suggests that the semi-analytic model fails 
to reproduce the evolution of galaxy population in the lowest density environments, overestimating
the level of aging. This suggests that the model should include
some physical process to stop the reddening of the group/cluster galaxy population 
in more isolated environments as was discussed previously.

So far, our analysis of SDSS groups proves that at a given concentration,  the degree of isolation for 
high mass groups may play a role at not only determining the number of group members but also at modulating the  
star formation histories (i.e. high mass, isolated groups have a smaller fraction of red galaxies).
Again, even though the mock catalogues are able to reproduce the dependence of number of members on isolation,
they fail to show the observed trend for the red galaxy fraction.

With the purpose of confirming that the galaxy population is intrinsically different
for groups of fixed mass and different degrees of concentration and isolation, we estimate
the magnitude difference between the brightest and second brightest group galaxies, $\Delta M_{12}$.
As it can be seen in Fig. ~\ref{fig:m12}, both the observed and simulated groups
 show a clear albeit weak tendency for higher values of $\Delta M_{12}$ for more concentrated 
groups (greater than $1\sigma$ detections, Tables 1 and 2), 
indicative of older group ages (Mendes de Oliveira \& Carrasco, 
2007). In the case of the isolation, as shown in the right panel of Fig. ~\ref{fig:m12}
 the dependence of $\Delta M_{12}$
is similar, although not that clear for massive SDSS groups.
These findings suggest that at a given mass, groups show
slightly different galaxy populations depending on their concentration and isolation.
It is important to recall that for the more massive systems, concentration and isolation are 
not correlated. 
Notice the higher typical values of $\Delta M_{12}$ in the mock catalogues  suggesting a larger
difference between central and satellite galaxies in the model than those in the SDSS.  This 
should be considered a possible problem in the galaxy formation modeling which suggests
again an overestimation  of the effects of environment in the semi-analytic scheme.

Finally, we studied the variation of group projected shapes on concentration and therefore group age.
We were not able to detect any significant variations in this sense for either the mock or
SDSS group samples, even though it is expected
that older groups or dark-matter haloes show rounder intrinsic and apparent shapes
(Paz et al., 2006).  Note that the analysis was carried out using the same narrow ($0.4$ dex)
ranges of group mass and different concentrations, that according to  our mock catalogues, should
have included groups with ages differing by up to $800$Myr.

\section{Discussion}

The results found in the previous sections on the effects of isolation and concentration on the
properties of groups and galaxy groups, can be used to shed light on several aspects
of the galaxy formation process, in particular in view of several discrepancies 
in the behaviour of real and simulated galaxies.  

We start by reviewing the results obtained for the mock catalogues.  In the mocks, more
isolated and concentrated groups, independently of their masses, show almost constant mass-to-light ratios, 
 slightly higher $\Delta M_{12}$   and higher red galaxy fractions. With respect to
the number of members, mock catalogues show different behaviours; low mass groups have
constant galaxy numbers independently of their concentration or isolation, and  massive groups show
opposing trends of increasing and decreasing number of members 
as the concentration and isolation increase (i.e. as the age increases), respectively.

On the other hand, high mass SDSS-DR6 groups show different trends depending on their
concentration and isolation.  We now analyse high- and low-mass group results
separately.  i) For high masses, the most isolated groups are characterised by higher
mass-to-light ratios, lower number of members and lower red galaxy fractions as a consequence
of being more populated by  late type galaxies. However, as a function of  increasing  concentration 
we find lower mass-to-light ratios and higher number of members 
and red galaxy fractions, in agreement with their older stellar populations as determined
by their spectral types.  
Considering that the group age is proportional to the concentration and anti-correlates
with the isolation, the discrepant trends of $M/L$ and red galaxy
fractions in SDSS-DR6 groups suggest different assembly processes indicated by different degrees of concentration
and isolation; in particular, the conditions placed by the large-scale structure around groups
seems to be important in shaping a group galaxy population such that more isolated groups
present lower numbers of galaxies, bluer colours and consequently larger mass-to-light ratios.
ii) For low mass SDSS groups, trends are more difficult to detect with either concentration
or isolation, except for a consistent decrease of $M/L$ with concentration and isolation (and
consistent with the results from the mocks).

As can be seen, the characteristics of galaxy groups in the SAM only reproduce the observations
to some level; in particular, massive groups show a dependence of $M/L$ and red galaxy fraction
 on isolation and concentration in stark contrast to the observational results.  
Also, the values of $\Delta M_{12}$ and red galaxy fractions are   higher than in the SDSS regardless
of group mass.
These problems may be understood by considering that more isolated groups are older in the model,
in contradiction with the results from the 
observations. This suggests an overestimation of the environmental effects on the evolution of systems,
with an excess of old galaxies in the lowest density environments. Particularly,
the treatment of satellites in the SAM, which lose their hot gas component and, consequently,
their sources of gas accretion when entering into a new dark-matter halo, could
be responsible for most of the trends observed in the mock groups, 
including the high values of red galaxy fraction and  $\Delta M_{12}$, as well as
the lower $M/L$ values.

The process thought to be responsible for the removal of cold gas from satellites 
is ram-pressure (Haynes \& Giovanelli, 1986, Solanes et al., 2001, 
and more recently Brueggen \& De Lucia, 2007, and references therein),
which some of the current models of semi-analytic galaxy formation (Cole et al., 2000, Baugh et al., 2005,
De Lucia et al., 2006, Croton et al., 2006, Bower et al., 2006, Lagos, Cora \& Padilla, 2008)
consider to produce an instantaneous result.  
Our observational results show several details that could be used to improve the modeling of processes that
drive the evolution of galaxies. For instance, if satellites in isolated groups or clusters were allowed 
to form stars after being accreted, the fractions of red galaxies and  $\Delta M_{12}$  values would decrease and the agreement with the observations would improve.

\section{Conclusions}

We have searched for possible dependencies of the galaxy population in groups on their assembly history, 
focusing on possible observational indicators of group age, with the aim of applying the study to the
SDSS-DR6 dataset.

In order to find indicators of group age, we use mock SDSS group  catalogues 
and find two suitable candidates: i) the  isolation of the group defined as the
distance to the nearest neighbor in terms of its virial radius, 
and ii) the concentration, measured as the density calculated using the fifth closest, bright galaxy
to the group centre.  Groups within narrow ranges of group mass in the mock catalogue show an increase of group
age with isolation and concentration.  We also find that the group age is correlated with the luminosity and colour
of group galaxies,  and with the magnitude difference between the brightest and second
brightest group galaxies.  

In order to test if these two theoretical age indicators can also be used in SDSS groups, 
we analyse the dependence of group spectral types on isolation and concentration finding consistent results 
only for the latter

Using both, the isolation and concentration of galaxy 
groups as indicators of their assembly history, we study properties of group galaxies
in SDSS and mock groups of similar mass. We find important variations in mass-to-light ratios, red galaxy fractions,
and  $\Delta M_{12}$.  These correlations or relations are not compatible between
model and observations, a problem that can be explained in terms of the differing trends found 
for age and spectral type as a function of isolation. We discuss how the modeling of the cold gas in
satellite galaxies in the SAM could be responsible for these discrepancies.

In this paper we have confirmed that in addition to a difference in clustering amplitude, groups
of similar mass and different assembly histories also show important differences in the characteristics of their
member galaxies, an aspect that can be added to current halo model and conditional luminosity
functions, and that can also be used to improve the modeling and understanding of the evolutionary
processes that shape galaxies in the observed Universe.

\section*{Acknowledgments}
We thank the Referee for his/her useful comments which have contributed to improve this paper.
This work was supported in part by the FONDAP ``Centro de Astrof\'\i sica", Fundaci\'on
Andes, Consejo Nacional de Ciencia y Tecnolog\'\i a (PIP 6446) y Agencia Nacional de Promoci\'on Cient\'\i fica y 
T\'ecnica (PICT 32342).  NP was supported by a Proyecto Fondecyt Regular No. 1071006.
This work was supported by the European Commission's ALFA-II programme through its funding of 
the Latin-american European Network for Astrophysics and Cosmology (LENAC).

\bsp

\label{lastpage}

\end{document}